\documentstyle[seceq,preprint]{ptptex}

\notypesetlogo

\title{%        %You can use \\ for explicit line-break
Stokes Theorem for Loop Variables of Non-Abelian Gauge Field
}
\author{%       %Use \sc for the family name
Minoru {\sc Hirayama}\footnote{hirayama@sci.toyama-u.ac.jp}
and Shugo {\sc Matsubara}
}

\inst{%         %Affiliation, neglected when [addenda] or [errata]
Department of Physics, Toyama University\\ Toyama 930-8555, JAPAN
}

\recdate{%      %Editorial Office will fill in this.
\today
}

\abst{%       %this abstract is neglected when [addenda] or [errata]
A simple analytic proof of the formula known as the non-Abelian Stokes theorem
is given. It is explicitly shown that the validity of the formula is 
guaranteed by the Bianchi identity for the gauge field. An attempt is made
to construct the Lagrangian for the gauge field in terms of loop variables.
}

\begin{document}

\maketitle
\section{Introduction}

The importance of the loop variable\\
\\
(A)\hspace{35mm}
$Pe^{ig \oint _\gamma A_\mu (x)dx^\mu} ,\hspace{3mm}$
$P$\hspace{1mm}:\hspace{1mm}path ordering\\
\\
has been stressed by many authors.
Here $A_\mu (x)$ is the non-Abelian gauge potential 
at a point $x$ lying on a closed unknotted loop $\gamma$ 
in the four-dimensional
Minkowskian space-time. 
As is well known, Wilson$^{1)}$ described the criterion 
of quark confinement by making use of the loop variable. 
According to the analysis of Wu and Yang,$^{2)}$ 
the field strength under-describes electromagnetism, 
but the loop integral of the gauge potential over-describes it. 
They discussed, however, that the Abelian version of (A) 
provides a complete description that is neither too much nor too little. 
They also discussed the role played by (A) in non-Abelian cases. 
Furthermore, it was suggested by Yang,$^{3)}$ Polyakov$^{4)}$
and Chan et al.$^{5)}$ that the non-Abelian gauge field theory 
might be formulated solely in terms of loop variables.

On the other hand, there exists a theorem,$^{6)-10)}$
the non-Abelian Stokes theorem (NAST), 
which equates the loop variable (A) with the quantity\\
\\
(B)\hspace{35mm}
${\cal P}e^{igI[S]}$,\hspace{5mm}
$I[S]=\int\!\!\!\int_S w(x)F_{\mu \nu}(x)w^{-1}(x)d\sigma^{\mu \nu}$,\\
\\
where ${\cal P}$ is a certain ordering operation, $w(x)$ is
an $x$-dependent unitary matrix, $S$ is a surface with the boundary 
$\partial S$ equal to $\gamma$, 
$d\sigma^{\mu \nu}$ is a surface element of $S$, 
and $F_{\mu \nu}(x)$ is the field strength given by
\begin{eqnarray}
   F_{\mu \nu}(x)= \partial_{\mu} A_{\nu}(x)-\partial_{\nu} A_{\mu}(x)
   -ig[A_{\mu}(x),A_{\nu}(x)].
\end{eqnarray}
It should be mentioned that some authors,$^{11),12)}$ with the help
of path integral, have replaced (B) by expressions
which do not contain the ordering operation ${\cal P}$.
It should also be noted that the NAST originally proposed by Halpern
in a special gauge takes a simpler form than (B).$^{13)}$

In our opinion, however, 
the proofs of the NAST proposed so far do not seem  satisfactory. 
The proof of the NAST given in Ref. 6) is 
complete but somewhat complicated. 
The discussion given in Ref. 7) is rather heuristic. 
The proofs in Refs. 8) and 9) are simple, 
but they suffer from some restrictions in the parametrization of the loop. 
These restrictions might cause inconveniences when we attempt 
to formulate a non-Abelian gauge theory in terms of the loop variable.
Hence these restrictions should be removed if possible. 
The first purpose of this article is to present a simple analytic proof
of the NAST without making use of these restrictions. 

The NAST asserts that (B) is equal to (A). 
\hspace{1mm}Although (A) depends only on $\gamma$\\$=\partial S$, 
the quantity (B) is defined by the integral over the surface $S$.
\hspace{1mm}It is then desirable to show explicitly that (B) is independent 
of the choice of $S$ if its boundary is fixed. 
The discussion given to this time regarding
this problem is also unsatisfactory. 
The authors of Ref. 7) considered the integral 
$\int\!\!\int_{S_1 +S_2} w(x)F_{\mu \nu}(x)w^{-1}(x)d\sigma^{\mu \nu}$, 
where $S_1 +S_2$ is a closed surface and $\partial S_1 =\gamma$, 
$\partial S_2 =\overline{\gamma}$  ($\gamma$ with the orientation reversed). 
They claimed that the integral vanishes if $F_{\mu \nu}(x)$ obeys 
the Bianchi identity
\begin{eqnarray}
   [D_\rho ,F_{\mu \nu}(x)]+[D_\mu ,F_{\nu \rho}(x)]
   +[D_\nu ,F_{\rho \mu}(x)]=0,\hspace{4mm}
   D_\mu = \partial_{\mu}-ig A_{\mu}(x). \nonumber \\
\end{eqnarray}
They then concluded that the surface integral $I[S]$ in (B) is 
independent of the choice of $S$ if $\partial S$ is fixed. 
The second purpose of this paper is to show that this is not the case. 
We shall show that $I[S]$ $does$ vary under deformations of $S$ 
with $\partial S$ fixed, but the quantity (B) nevertheless remains 
fixed if the Bianchi identity is satisfied. 
We thus establish the equality (A)=(B) 
and find that the surface $S$ in (B) may be 
arbitrary as long as it satisfies $\partial S$ = $\gamma$.
We shall also find another important role played by the Bianchi identity: 
it insures the commutativity of differentiations of a loop variable
in parameters which specify the loop.
We note that some properties implied by the surface independence
of (B) were discussed in Ref. 10).

Some time ago, an attempt was made to construct the action of the non-Abelian
gauge field in terms of loop variables.$^{5)}$ It seems, however, 
that the averaging procedure adopted there is somewhat ambiguous. 
The third purpose of this paper is to achieve the same attempt 
as that of Ref. 5)
without ambiguity.

This paper is organized as follows. 
In $\S 2$, we describe a simple analytic proof of NAST 
by generalizing the method of Brali\'{c}. 
In $\S 3$, we calculate the variation $\delta I[S]$ 
under a small deformation of $S$ with $\partial S$ fixed.
We shall see that it does not vanish even if the Bianchi identity 
is satisfied. In $\S 4$, we proceed to the calculation 
of the variation $\delta ({\cal P}e^{igI[S]})$ 
under the same deformation of $S$.
We find that it vanishes if $F_{\mu \nu}(x)$ satisfies 
the Bianchi identity. 
Another relation between the loop variable and the Bianchi identity is
explored in $\S 5$. 
We find that the commutativity of differentaions of the loop or 
string variable
with respect to parameters specifying it requires the Bianchi identity 
for $F_{\mu \nu}(x)$.
In $\S 6$, we attempt to express the action of the non-Abelian gauge field 
in terms of loop variables.
The final section, $\S 7$, is devoted to summary.

\section{Simple analytic proof of NAST}

 Brali\'{c}'s proof$^{8)}$ of the NAST is analytic and seems to be simpler
than Aref'eva's diagramatic proof,$^{6)}$
which makes use of infinite products.
In this section, we slightly generalize Brali\'{c}'s discussion, 
remove an unnecessary assumption he made, 
and obtain the desired form of the NAST.

Suppose that a point in the four-dimensional Minkowski space $M$ is specified
by differentiable functions $x(\kappa)=
(x^0 (\kappa),x^1 (\kappa),x^2 (\kappa),x^3 (\kappa))$
of four real parameters $\kappa =(s,t,u,v)$. 
An oriented string in $M$ is given by, 
e.g., $\{ x(\kappa) |t,u,v:$ fixed$, \hspace{1mm}s:s_1 $\\$\rightarrow s_2 \}$,
where $s:s_1\rightarrow s_2$ means that $s$ varies from $s_1$ to $s_2$.
Similarly an oriented loop in $M$ is given by, 
e.g., $\{ x(\kappa) |u,v:$ fixed$, \hspace{1mm}(s,t)\in \sigma \}$
with $\sigma$ an oriented closed loop in the $(s,t)$-plane.
To keep up the one-to-one correspondence between $x(\kappa)$ and $\kappa$,
the Jacobian of the mapping should not vanish for any $\kappa$:
\begin{eqnarray}
   \frac{\partial (x^0 (\kappa),x^1 (\kappa),x^2 (\kappa),x^3 (\kappa))}
   {\partial (s,t,u,v)}\neq 0.
\end{eqnarray}
The importance of this condition will be discussed again in $\S 6$.

We first consider the string variable $U(s_2,s_1;t)$ associated with the string
$\{x(\kappa) |t,u,v \\
:$ fixed$, \hspace{1mm}s:s_1\rightarrow s_2 \}$ :
\begin{eqnarray}
   U(s_2,s_1;t)=Pe^{ig \int_{s_1}^{s_2} A_\mu (x(s,t))x^\mu_s(s,t)ds}.
\end{eqnarray}
Here $P$ denotes the path ordering, the fixed parameters $u$ and $v$
are suppressed, and $x^\mu_s$ is defined by
\begin{eqnarray}
   x^\mu_s(s,t)=\frac{\partial x^\mu(s,t)}{\partial s}.
\end{eqnarray}
Then we have
\begin{eqnarray}
   U(s_2,s;t)U(s,s_1;t)= U(s_2,s_1;t)
\end{eqnarray}
and
\begin{eqnarray}
   \frac{U(s_2,s_1;t)}{\partial s_2}&=&j_s(s_2,t)U(s_2,s_1;t),\\
   \frac{U(s_2,s_1;t)}{\partial s_1}&=&-U(s_2,s_1;t)j_s(s_1,t),
\end{eqnarray}
where $j_s(s,t)$ is given by
\begin{eqnarray}
   j_s(s,t)=igA_\mu(x(s,t))x^\mu_s(s,t).
\end{eqnarray}
To calculate the $t$-derivative of $U(s_2,s_1;t)$, 
we make use of Brali\'{c}'s formula,$^{8)}$
\begin{eqnarray}
\frac{\partial}{\partial s}\left\{ U(s_2,s;t)\left(\frac{\partial}{\partial t}
-j_t(s,t)\right)U(s,s_1;t)\right\}=-U(s_2,s;t)K(s,t)U(s,s_1;t), \nonumber \\
\end{eqnarray}
with $j_t(s,t)$ and $K(s,t)$ defined by
\begin{eqnarray}
j_t(s,t)&=&igA_\mu(x(s,t))x^\mu_t(s,t),
\hspace{5mm}x^\mu_t(s,t)=\frac{\partial x^\mu(s,t)}{\partial t},\\
K(s,t)&=&igF_{\mu \nu}(x(s,t))x^\mu_s(s,t)x^\nu_t(s,t).
\end{eqnarray}
Integrating (2$\cdot$8) with respect to $s$ from $s_1$ to $s_2$ 
and making use of (2$\cdot$4),
we obtain
\begin{eqnarray}
\frac{\partial U(s_2,s_1;t)}{\partial t}U^{-1}(s_2,s_1;t)
=&-&\int_{s_1}^{s_2}ds \hspace{1mm}
U(s_2,s;t)K(s,t)U^{-1}(s_2,s;t) \nonumber \\
&+&j_t(s_2,t)-U(s_2,s_1;t)j_t(s_1,t)U^{-1}(s_2,s_1;t). \nonumber \\
\end{eqnarray}
Brali\'{c} assumes that the parameters $s$ and $t$
and the functions $x^\mu(s,t)$, $\mu$ = 0, 1, 2, 3, are chosen such that
$x^\mu_t(s,t)$ vanishes at $s$ = $s_1$  and $s$ = $s_2$. Then $j_t(s_2,t)$
and $j_t(s_1,t)$ on the r.h.s.\hspace{1mm}of (2$\cdot$11) vanish. 
Although this assumption  
considerably simplifies the formula for 
$\frac{\partial U(s_2,s_1;t)}{\partial t}$, we do not adopt it because
it violates (2$\cdot$1). Instead we proceed in the following way.

We define another string variable, $V(t_2,t_1;s)$, by 
\begin{eqnarray}
V(t_2,t_1;s)=Pe^{ig \int_{t_1}^{t_2} A_\mu (x(s,t))x^\mu_t(s,t)dt}
\end{eqnarray}
corresponding to the string $\{x(\kappa)|
s,u,v:$ fixed$,\hspace{1mm}t:t_1\rightarrow t_2\}$.
In analogy to (2$\cdot$4)-(2$\cdot$6) and (2$\cdot$11) 
for $U(s_2,s_1;t)$, we obtain
\begin{eqnarray}
V(t_2,t;s)V(t,t_1;s)&=&V(t_2,t_1;s),\\
\frac{V(t_2,t_1;s)}{\partial t_2}&=&j_t(s,t_2)U(t_2,t_1;s),\\
\frac{V(t_2,t_1;s)}{\partial t_1}&=&-V(t_2,t_1;s)j_t(s,t_1)
\end{eqnarray}
and
\begin{eqnarray}
\frac{\partial V(t_2,t_1;s)}{\partial s}V^{-1}(t_2,t_1;s)
=\int_{t_1}^{t_2}dt \hspace{1mm} &V(&t_2,t;s)K(s,t)V^{-1}(t_2,t;s) \nonumber \\
+j_s(s,&t_2)&-V(t_2,t_1;s)j_s(s,t_1)V^{-1}(t_2,t_1;s). \nonumber \\
\end{eqnarray}
We now consider a loop variable given as a product of the $U$ and the $V$:
\begin{eqnarray}
W(s,t)=V(t_1,t;s_1)U(s_1,s;t)V(t,t_1;s)U(s,s_1;t_1).
\end{eqnarray}
The loop corresponding to $W(s,t)$ is given by
$\{x(s',t',u,v)|u,v:$ fixed$,\hspace{1mm}(s',t')\in\sigma'\}$,
where $\sigma '$ is the counterclockwise boundary of the rectangle 
$\{(s',t')|s_1\leq s'\leq s,\hspace{1mm} t_1 $\\$\leq t'\leq t\}$
starting and ending at $(s_1,t_1)$.
With the help of the above formulas and the properties  
$\frac{\partial V(t_1,t;s_1)}{\partial s}$= 0 and 
$\frac{\partial U(s,s_1;t_1)}{\partial t}$= 0, it is straightforward
to calculate $s$- and $t$-derivatives of $W(s,t)$. It can be seen that
terms containing $j_t(s,t)$ or $j_s(s,t)$ cancel out in
$\frac{\partial W(s,t)}{\partial s}$ and 
$\frac{\partial W(s,t)}{\partial t}$.
Setting $t=t_2$ in $\frac{\partial W(s,t)}{\partial s}$ and 
$s=s_2$ in $\frac{\partial W(s,t)}{\partial t}$,
we obtain the simple results
\begin{eqnarray}
\frac{\partial W(s,t_2)}{\partial s}W^{-1}(s,t_2)
=\int_{t_1}^{t_2}dt \hspace{1mm} u(s,t)K(s,t)u^{-1}(s,t),\\
\frac{\partial W(s_2,t)}{\partial t}W^{-1}(s_2,t)
=\int_{s_1}^{s_2}ds \hspace{1mm} v(s,t)K(s,t)v^{-1}(s,t),
\end{eqnarray}
where $u(s,t)$ and $v(s,t)$ are given by
\begin{eqnarray}
u(s,t)&=&V(t_1,t_2;s_1)U(s_1,s;t_2)V(t_2,t;s),\\
v(s,t)&=&V(t_1,t;s_1)U(s_1,s;t).
\end{eqnarray}
Integrating (2$\cdot$18) with respect to $s$ from $s_1$ to $s_2$ and recalling
the definition (2$\cdot$10), we arrive at the NAST,
\begin{eqnarray}
W[\gamma]&=&P_s \exp \left(\int_{s_1}^{s_2}ds\int_{t_1}^{t_2}dt\hspace{1mm}
u(s,t)K(s,t)u^{-1}(s,t)\right) \nonumber \\
&=&P_s \exp \left(ig\int_{s_1}^{s_2}ds\int_{t_1}^{t_2}dt \hspace{1mm}
u(s,t)F_{\mu\nu}(x(s,t))x^\mu_s(s,t)x^\nu_t(s,t)u^{-1}(s,t)\right),
 \nonumber \\
\end{eqnarray}
where $P_s$ is the $s$-ordering and $W[\gamma]$ is defined by
\begin{eqnarray}
W[\gamma]&=&Pe^{ig\oint_\gamma A_\mu(x)dx^\mu},\\
\gamma&=&\{x(s,t,u,v)|u,v:\hspace{1mm}fixed,\hspace{1mm} (s,t)\in\sigma \}
\end{eqnarray}
with $\sigma$ the counterclockwise boundary of the rectangle 
$\{(s,t)|s_1\leq s\leq s_2, \hspace{1mm}
t_1\leq t\leq t_2\}$ in the $(s,t)$-plane
starting and ending at the point $(s_1,t_1)$.
Similarly, we obtain
\begin{eqnarray}
W[\gamma]=P_t \exp \left(\int_{t_1}^{t_2}dt\int_{s_1}^{s_2}ds \hspace{1mm}
v(s,t)K(s,t)v^{-1}(s,t)\right)
\end{eqnarray}
from (2$\cdot$19), where $P_t$ is the $t$-ordering. We note that the NAST
 (2$\cdot$25) is the same as Aref'eva's NAST, while (2$\cdot$22)
is of a slightly different form.
It is clear that both (2$\cdot$22) and (2$\cdot$25) 
can be rewritten
in the form of (B) in $\S$1.

\section{Dependence of $I[S]$ on $S$}

For the loop variable $W[\gamma]$ defined by (2$\cdot$23) and (2$\cdot$24),
we have obtained the formula
\begin{eqnarray}
W[\gamma]&=&P_te^{igI[S]},\\
igI[S]&=&\int_{t_1}^{t_2}dt\int_{s_1}^{s_2}ds \hspace{1mm}
v(s,t)K(s,t)v^{-1}(s,t)
\end{eqnarray}
with $v(s,t)$, $K(s,t)$ and $S$ given by (2$\cdot$21), (2$\cdot$10) and
$S=\{x(s,t,u,v)|u,v:$ fixed$,\\
s_1\leq s\leq s_2,\hspace{1mm}t_1\leq t\leq t_2\}$,
respectively.
The authors of Ref. 7)  
claimed that $I[S_1+S_2]$ vanishes if 
$S_1+S_2$ is a closed surface and concluded that $I[S]$ 
is independent of the choice of $S$ provided 
that its boundary $\partial S$ is fixed.

In this section, we show by explicit calculation 
that this is not the case: $I[S]$ varies 
under deformations of $S$ with $\partial S$ fixed.
We first discuss why $I[S_1+S_2]$ cannot be expected to vanish.
Suppose that two surfaces $S_1$ and $S_2$ satisfy $\partial S_1=\gamma$ and
$\partial S_2=\overline{\gamma}$.
Then we have $W[\gamma]=P_te^{igI[S_1]}$ and 
$W[\overline{\gamma}]=P_te^{igI[S_2]}$.
Since the product $W[\gamma]W[\overline{\gamma}]$ is equal to 1, we are led to
\begin{eqnarray}
(P_te^{igI[S_1]})(P_te^{igI[S_2]})=1.
\end{eqnarray}
We, however, find no reason based on (3$\cdot$3) 
to expect $I[S_1+S_2]$ to vanish.

We now proceed to the calculation of the variation $\delta I[S]$
under a small deformation of $S$ with $\partial S=\gamma$ fixed. 
Such a deformation of $S$ is realized by the variation 
$x(s,t)\rightarrow x(s,t)+\delta x(s,t)$ with $\delta x(s,t)$
satisfying
\begin{eqnarray}
\delta x(s,t)=0;\hspace{2mm}s=s_1\hspace{1mm}or\hspace{1mm}
 s_2\hspace{2mm}
 and/or\hspace{2mm} t=t_1\hspace{1mm}or\hspace{1mm} t_2.
\end{eqnarray}
The variation of $I[S]$ is calculated to be 
\begin{eqnarray}
\delta I[S]&=&\int_{t_1}^{t_2}dt\int_{s_1}^{s_2}ds \hspace{1mm}
\delta x^\rho(s,t)\frac{\delta I[S]}{\delta x^\rho(s,t)},\\
ig\frac{\delta I[S]}{\delta x^\rho(s,t)}&=&
\int_{t_1}^{t_2}dt'\int_{s_1}^{s_2}ds' \hspace{1mm}
v'\left(\frac{\delta K'}{\delta x^\rho(s,t)}
+[v'^{-1}\frac{\delta v'}{\delta x^\rho(s,t)},K']\right) v'^{-1},
\end{eqnarray}
with abbreviated notation $v'=v(s',t')$, $K'=K(s',t')$.
Through a rather tedious calculation (see Appendix A), we obtain
\begin{eqnarray}
\delta I[S]&=&\int_{t_1}^{t_2}dt\int_{s_1}^{s_2}ds \hspace{1mm}
\delta x^\rho(s,t)\{E_\rho(s,t)+G_\rho(s,t)\},\\
E_\rho(s,t)&=&v\left\{[D_\rho,F_{\mu \nu}(x)]+[D_\mu,F_{\nu \rho}(x)]
+[D_\nu,F_{\rho \mu}(x)]\right\}v^{-1}x^\mu_s x^\nu_t , \\
G_\rho(s,t)&=&\biggl[v(s,t)K_\rho(s,t)v^{-1}(s,t),
\int_{s_1}^{s_2}ds' \hspace{1mm}
v(s',t)K(s',t)v^{-1}(s',t)\biggl],
\end{eqnarray}
where $v=v(s,t)$, $F_{\mu \nu}(x)=F_{\mu \nu}(x(s,t))$,
$x^\mu_s=x^\mu_s(s,t)$, $x^\mu_t =x^\mu_t(s,t)$ and $K_\rho(s,t)$ is defined by
\begin{eqnarray}
K_\rho(s,t)=F_{\mu \rho}(x(s,t))x^\mu_s(s,t).
\end{eqnarray}
The Bianchi identity (1$\cdot$2) indicates that $E_\rho(s,t)$ vanishes.
We then see that the r.h.s.of (3$\cdot$7) vanishes 
only if $\delta x^\rho(s,t)$ is
of the form $x^\rho_s(s,t)\delta f(t)+x^\rho_t(s,t)\delta g(t)$,
where $\delta f(t)$ and $\delta g(t)$ are $s$-independent functions.
We conclude that $\delta I[S]$ does not vanish for general deformations 
of $S$ satisfying (3$\cdot$4). We shall show in the next section, however,
that the variation $\delta(P_te^{igI[S]})$ nonetheless vanishes.

\section{Independence of $P_te^{igI[S]}$ on the choice of $S$}

In this section, we show that the variation of $P_te^{igI[S]}$
under the deformation of $S$ satisfying (3$\cdot$4) is given by
\begin{eqnarray}
\delta (P_te^{igI[S]})=
ig\int_{s_1}^{s_2}ds\int_{t_1}^{t_2}dt \hspace{1mm}
\delta x^\rho(s,t)X(t)E_\rho(s,t)Y(t),
\end{eqnarray}
where $E_\rho(s,t)$ is defined in (3$\cdot$8) 
and $X(t)$ and $Y(t)$ are given by
\begin{eqnarray}
X(t)&=&P_te^{\int_{t}^{t_2}B(t')dt'},\\
Y(t)&=&P_te^{\int_{t_1}^{t}B(t')dt'}=X^{-1}(t)W[\gamma],\\
B(t)&=&\int_{s_1}^{s_2}ds \hspace{1mm} v(s,t)K(s,t)v^{-1}(s,t).
\end{eqnarray}
Here the functions $v(s,t)$ and $K(s,t)$ and the surface $S$ are
those defined in the previous section. The proof of (4$\cdot$1) is
given in the following way.

Noting that $I[S]$ is given as
\begin{eqnarray}
igI[S]=\int_{t_1}^{t_2}B(t)dt, \nonumber
\end{eqnarray}
we have 
\begin{eqnarray}
\delta (P_te^{igI[S]})=\int_{t_1}^{t_2}dtX(t)\delta B(t)Y(t).
\end{eqnarray}
Since the variation $\delta x^\rho(s,t)$ vanishes 
on the boundary of $S$, we have
\begin{eqnarray}
\delta B(t)&=&\delta\left(\int_{s_1}^{s_2}ds \hspace{1mm}
v(s,t)K(s,t)v^{-1}(s,t)\right) \nonumber \\
&=&V(t_1,t;s_1)\left[\int_{s_1}^{s_2}ds \hspace{1mm}
\delta\{U(s_1,s;t)K(s,t)U^{-1}(s_1,s;t)\}\right]
V^{-1}(t_1,t;s_1).\nonumber \\
\end{eqnarray}
Similar to the case of the calculation in the previous section, we obtain
\begin{eqnarray}
\delta\{U(s_1,s;t)K(&s,&t)U^{-1}(s_1,s;t)\} \nonumber \\
=\int_{s_1}^{s_2}&ds'&\int_{t_1}^{t_2}dt'  \hspace{1mm}
\delta x^\rho(s',t')U(s_1,s;t) \nonumber \\
&\times& \left(\frac{\delta K(s,t)}{\delta x^\rho(s',t')}
+[U^{-1}(s_1,s;t)\frac{\delta U(s_1,s;t)}{\delta x^\rho(s',t')}
,K(s,t)]\right) U^{-1}(s_1,s;t).\nonumber \\
\end{eqnarray}
From the above, we are led to
\begin{eqnarray}
\delta (P_te^{igI[S]})=P+Q,
\end{eqnarray}
where $P$ and $Q$ are given by
\begin{eqnarray}
P=\int_{s_1}^{s_2}&ds&\int_{t_1}^{t_2}dt \hspace{1mm} X(t)v(s,t) \nonumber \\
&\times& \left(\int_{s_1}^{s_2}ds'\int_{t_1}^{t_2}dt' \hspace{1mm} 
\delta x^\rho(s',t')
\frac{\delta K(s,t)}{\delta x^\rho(s',t')}\right)v^{-1}(s,t)Y(t),\\
Q=\int_{s_1}^{s_2}&ds&\int_{t_1}^{t_2}dt  \hspace{1mm} X(t)v(s,t) \nonumber \\
&\times& \left(\int_{s_1}^{s_2}ds'\int_{t_1}^{t_2}dt' \hspace{1mm} 
\delta x^\rho(s',t')
\biggl[U^{-1}(s_1,s;t)
\frac{\delta U(s_1,s;t)}{\delta x^\rho(s',t')},
K(s,t)\biggl]\right) \nonumber \\
&\times& v^{-1}(s,t)Y(t). 
\end{eqnarray}
In Appendices B and C, we describe the details of the calculation of $P$
and $Q$, respectively. It turns out that they are given by
\begin{eqnarray}
P&=&P_{21}+P_{22}-R,\\
Q&=&-P_{22}+R,
\end{eqnarray}
where $P_{21}$, $P_{22}$ and R are given by (B$\cdot$9), (B$\cdot$10) 
and (B$\cdot$8), respectively. We thus obtain the simple result (4$\cdot$1).
From this formula, we see that the $\delta (P_te^{igI[S]})$ vanishes, 
as it should if the Bianchi identity (1$\cdot$2) is 
imposed on $F_{\mu \nu}(x)$.

\section{Another role of the Bianchi identity}

In the previous sections, 
we have given a complete proof of the NAST and observed 
that the validity of the NAST is guaranteed by the Bianchi identity.
In this section we show that
the Bianchi identity insures the commutativity of
differentiations of a loop variable with respect to parameters
specifying the loop.
In $\S$1, we parametrized the loop $\gamma$ by four
parameters $\kappa =(s,t,u,v)$. For the discussion below, 
it is convenient to parametrize $\gamma$
in the following way:
\begin{eqnarray}
\gamma = \{x(r,u,v)|u,v:\hspace{1mm}fixed, \hspace{1mm} r:0\rightarrow 4\}.
\end{eqnarray}
Here $r$ corresponds to the pair $(s,t)$ on $\gamma$ and the portions
$\{(s,t)|t=t_1, \hspace{1mm}s:s_1\rightarrow s_2\}$, 
$\{(s,t)|s=s_2, \hspace{1mm}t:t_1\rightarrow t_2\}$, 
$\{(s,t)|t=t_2, \hspace{1mm}s:s_2\rightarrow s_1\}$ and
$\{(s,t)|s=s_1, \hspace{1mm}t:t_2\rightarrow t_1\}$ 
correspond to the intervals
$\{r:0\rightarrow 1\}$,
$\{r:1\rightarrow 2\}$,
$\{r:2\rightarrow 3\}$ and
$\{r:3\rightarrow 4\}$, respectively.
 A calculation similar to that leading us to
(2$\cdot$11) yields
\begin{eqnarray}
\frac{\partial W[\gamma]}{\partial v}=
-\int_0^4dr \hspace{1mm} 
\omega(&4,r&)K_v(r)\omega(r,0)+j_v(0)W[\gamma]-W[\gamma]j_v(0),\\
\omega(r_2,r_1)&=&P_re^{ig\int_{r_1}^{r_2}A_\mu(x(r'))x_r^\mu(r')dr'},\\
j_v(r)&=&igA_\mu(x(r))x_v^\mu(r),\\
K_v(r)&=&igF_{\mu \nu}(x(r))x_r^\mu(r)x_v^\nu(r),\\
x_r^\mu(r)&=&\frac{\partial x^\mu(r)}{\partial r},\hspace{3mm}
x_v^\mu(r)=\frac{\partial x^\mu(r)}{\partial v},
\end{eqnarray}
where parameters $u$ and $v$ are suppressed and the pair $(s,t)$ is 
replaced by a single parameter $r$.
Differentiating (5$\cdot$2) with respect to $u$, we have
\begin{eqnarray}
\frac{\partial}{\partial u}\left(
\frac{\partial W[\gamma]}{\partial v}\right)
&=&-\int_0^4dr\biggl\{\frac{\partial \omega(4,r)}{\partial u}K_v(r)
\omega(r,0)
+\omega(4,r)\frac{\partial K_v(r)}{\partial u}\omega(r,0) \nonumber \\
& &\hspace{17mm}
+\omega(4,r)K_v(r)\frac{\partial \omega(r,0)}{\partial u}\biggl\} \nonumber \\
& &+\left[\frac{\partial j_v(0)}{\partial u},W[\gamma]\right]
+\left[j_v(0) ,\frac{\partial W[\gamma]}{\partial u}\right] \nonumber \\
&=&S_{uv}+T_{uv},
\end{eqnarray}
where $S_{uv}$ and $T_{uv}$ are given by
\begin{eqnarray}
S_{uv}&=&\int_0^4dr\int_r^4dr' \hspace{1mm}
\omega(4,r')K_u(r')\omega(r',r)K_v(r)\omega(r,0) \nonumber \\
& &+\int_0^4dr\int_0^rdr' \hspace{1mm}
\omega(4,r)K_v(r)\omega(r,r')K_u(r')\omega(r',0) \nonumber \\
&=&\int_0^4dr\int_r^4dr' \hspace{1mm}
\omega(4,r') \nonumber \\
& &\hspace{3mm}\times \left\{K_u(r')\omega(r',r)K_v(r)
+K_v(r')\omega(r',r)K_u(r)\right\}\omega(r,0),
\end{eqnarray}
\begin{eqnarray}
T_{uv}&=&\left[\frac{\partial j_v(0)}{\partial u},W[\gamma]\right]
+\left[j_v(0) ,\frac{\partial W[\gamma]}{\partial u}\right] \nonumber \\
& &\hspace{1mm}-\int_0^4dr\hspace{1mm}
\biggl\{j_u(0)\omega(4,r)K_v(r)\omega(r,0)
-\omega(4,r)j_u(r)K_v(r)\omega(r,0)\nonumber \\
& &\hspace{2cm}+\omega(4,r)K_v(r)j_u(r)\omega(r,0)
-\omega(4,r)K_v(r)\omega(r,0)j_u(0)\nonumber \\
& &\hspace{2cm}
+\omega(4,r)\frac{\partial K_v(r)}{\partial u}\omega(r,0)\biggl\},\\
& &\hspace{10mm}j_u(r)=igA_\mu(x(r))x_u^\mu(r),\hspace{3mm} 
x_u^\mu(r)=\frac{\partial x^\mu (r)}{\partial u},\\
& &\hspace{10mm}K_u(r)=igF_{\mu \nu}(x(r))x_r^\mu(r)x_u^\nu(r).
\end{eqnarray}
Similarly, we obtain
\begin{eqnarray}
\frac{\partial }{\partial v}
\left(\frac{\partial W[\gamma]}{\partial u}\right)
=S_{vu}+T_{vu},
\end{eqnarray}
where the notation is self-evident. We easily see
\begin{eqnarray}
S_{uv}-S_{vu}=0.
\end{eqnarray}
After some calculations, we obtain (see Appendix D)
\begin{eqnarray}
T_{uv}-T_{vu}=ig\int_0^4&dr&  \hspace{1mm} 
x_r^\mu(r)x_u^\nu(r)x_v^\rho(r) \nonumber \\
&\times& \omega(4,r)
([D_\rho,F_{\mu \nu}(x)]+[D_\nu,F_{\rho \mu}(x)]
+[D_\mu,F_{\nu \rho}(x)])\omega(r,0). \nonumber \\
\end{eqnarray}
From
(5$\cdot$7), (5$\cdot$13)
and (5$\cdot$14), we are led to the result
\begin{eqnarray}
\frac{\partial }{\partial u}
\left(\frac{\partial W[\gamma]}{\partial v}\right)
&-&\frac{\partial }{\partial v}
\left(\frac{\partial W[\gamma]}{\partial u}\right)
 \nonumber \\
=ig\int_0^4&dr&  \hspace{1mm} 
x_r^\mu(r)x_u^\nu(r)x_v^\rho(r) \nonumber \\
&\times& \omega(4,r)
([D_\rho,F_{\mu \nu}(x)]+[D_\nu,F_{\rho \mu}(x)]
+[D_\mu,F_{\nu \rho}(x)]
)\omega(r,0).\nonumber \\
\end{eqnarray}
From this formula, we conclude that the commutativity of $u$- and 
$v$-derivatives on the loop variable 
$W[\gamma]$ is guaranteed by the Bianchi identity.

\section{Lagrangian density and the loop variable}

The Lagrangian density of the non-Abelian gauge field 
in the four dimensional Minkowski space is proportional to 
tr$\{F_{\mu \nu}(x)F^{\mu \nu}(x)\}$. In this section, we attempt to express 
the last quantity in terms of loop and/or string variables. We define
the parameters $\kappa ^\alpha$, $\alpha=0,1,2,3$, 
by $\kappa =(s,t,u,v) =(\kappa^0, \kappa^1, \kappa^2, \kappa^3)$.
Then, Eqs. (2$\cdot$18) and (2$\cdot$10) lead us to the equality
\begin{eqnarray}
\frac{\partial}{\partial \kappa^\beta}
\left[\frac{\partial W^{(\alpha \beta)}(\kappa)}{\partial \kappa^\alpha}
\{W^{(\alpha \beta)}(\kappa)\}^{-1}\right]&=&
ig w^{(\alpha \beta)}(\kappa)F_{\mu \nu}(x(\kappa))x^\mu_\alpha(\kappa)
x^\nu_\beta (\kappa)\{w^{(\alpha \beta)}(\kappa)\}^{-1},\nonumber\\
\\
x^\mu_\alpha(\kappa)&=&\frac{\partial x^\mu (\kappa)}{\partial \kappa^\alpha},
\end{eqnarray}
where $W^{(\alpha \beta)}(\kappa)$ ($w^{(\alpha \beta)}(\kappa)$) is 
a loop (string) variable defined by a loop (string) which passes (starts from)
the point $\kappa$ and lies in the 
$\kappa^\alpha \kappa^\beta $-plane. We note that parameters such as $s_1$
and $t_1$ in (2$\cdot$17) are suppressed here.
Recalling that the loop variable 
$W^{(\alpha \beta)}(\kappa)$ is defined by a counterclockwise loop,
we should set 
\begin{eqnarray}
W^{(\beta \alpha)}(\kappa)=\{W^{(\alpha \beta)}(\kappa)\}^{-1}.
\end{eqnarray}
since a counterclockwise loop in the $\kappa^\alpha \kappa^\beta $-plane
is a clockwise loop in the $\kappa^\beta \kappa^\alpha $-plane.
The consistency of (6$\cdot$1) with (6$\cdot$3), 
$\frac{\partial}{\partial \kappa^\alpha}
\frac{\partial}{\partial \kappa^\beta}
W^{(\alpha \beta)}(\kappa)=
\frac{\partial}{\partial \kappa^\beta}
\frac{\partial}{\partial \kappa^\alpha}
W^{(\alpha \beta)}(\kappa)$ and 
$F_{\mu \nu}(x(\kappa))=-F_{\nu \mu}(x(\kappa))$ requires
\begin{eqnarray}
w^{(\beta \alpha)}(\kappa)=
\{W^{(\alpha \beta)}(\kappa)\}^{-1}w^{(\alpha \beta)}(\kappa).
\end{eqnarray}
From (6$\cdot$1), we have
\begin{eqnarray}
x^\mu_\alpha (\kappa) x^\nu_\beta (\kappa) 
x^\rho_\gamma (\kappa)x^\sigma_\delta (\kappa)
tr\{F_{\mu \nu}(x(\kappa))F_{\rho \sigma}(x(\kappa))\}
=L_{\alpha \beta \gamma \delta}(\kappa),
\end{eqnarray}
where $L_{\alpha \beta \gamma \delta}(\kappa)$ is defined by
\begin{eqnarray}
L_{\alpha \beta \gamma \delta}(\kappa)
=\frac{1}{(ig)^2}&tr&\biggl[\{w^{(\alpha \beta)}(\kappa)\}^{-1}
\left\{\frac{\partial}{\partial \kappa^\beta}
\left(\frac{\partial W^{(\alpha \beta)}(\kappa)}{\partial \kappa^\alpha}
\{W^{(\alpha \beta)}(\kappa)\}^{-1}\right)\right\}
w^{(\alpha \beta)}(\kappa) \nonumber \\
& &\hspace{3mm}\times \{w^{(\gamma \delta)}(\kappa)\}^{-1}
\left\{\frac{\partial}{\partial \kappa^\delta}
\left(\frac{\partial W^{(\gamma \delta)}(\kappa)}{\partial \kappa^\gamma}
\{W^{(\gamma \delta)}(\kappa)\}^{-1}\right)\right\}
w^{(\gamma \delta)}(\kappa)\biggl]. \nonumber \\
\end{eqnarray}
Here we specify the parametrization of loops by
\begin{eqnarray}
g^{\alpha \beta}(\kappa)x^\mu_\alpha(\kappa)x^\nu_\beta(\kappa)=\eta^{\mu \nu},
\end{eqnarray}
where $g^{\alpha \beta}(\kappa)$ is the inverse 
of the metric tensor $g_{\alpha \beta}(\kappa)$ in 
the parameter space and $\eta^{\mu \nu}=$ diag$(1,-1,-1,-1)$ is 
the Minkowski metric.
The condition (6$\cdot$7) implies that $x^\mu_\alpha(\kappa)$ is 
now regarded as a tetrad satisfying (2$\cdot$1). 
Then we have $g_{\alpha \beta}(\kappa)d\kappa^\alpha d\kappa^\beta$\\
$=\eta_{\mu \nu}dx^\mu dx^\nu $, which implies that an infinitesimal distance
in the parameter space coincides with that in the space-time.
From (6$\cdot$5) and 
(6$\cdot$7), we obtain
\begin{eqnarray}
tr\{F_{\mu \nu}(x(\kappa))F^{\mu \nu}(x(\kappa))\}
=g^{\alpha \gamma}(\kappa)g^{\beta \delta}(\kappa)
L_{\alpha \beta \gamma \delta}(\kappa).
\end{eqnarray}
We have thus expressed
 $tr\{F_{\mu \nu}(x(\kappa))F^{\mu \nu}(x(\kappa))\}$
in terms of loop and string variables. 
We note that it is possible to remove string variables
by further specifying the parametrization of loops. For example, if we put
$\kappa^\alpha = x^\alpha$, we have
\begin{eqnarray}
tr\{F_{\mu \nu}(x)F^{\mu \nu}(x)\}
&=&\frac{1}{(ig)^2}tr\biggl[\left\{\frac{\partial}{\partial x^\nu}
\left(\frac{\partial W^{(\mu \nu)}(x)}{\partial x^\mu}
\{W^{(\mu \nu)}(x)\}^{-1}\right)\right\} \nonumber \\
& &\hspace{2cm}\times
\left\{\frac{\partial}{\partial x_\nu}
\left(\frac{\partial W^{(\mu \nu)}(x)}{\partial x_\mu}
\{W^{(\mu \nu)}(x)\}^{-1}\right)\right\}\biggl]. \nonumber \\
\end{eqnarray}
We observe that on the r.h.s. of (6$\cdot$9) 
there appear twelve loop variables
$W^{(\mu \nu)}(x)$, $\mu,\nu=0,1,2,3$, $\mu\neq\nu$, which correspond to 
rectangular loops in six $x^\mu x^\nu$-planes meeting at a vertex $x$.
Noting the relation (6$\cdot$3), we understand that the set of 
six loop variables $W^{(\mu \nu)}(x)$, $\mu >\nu$, describes 
the Lagrangian density of the non-Abelian gauge field.
Note that we could not impose a condition such as (6$\cdot$7)
if we adopted Brali\'{c}'s restriction in parametrizing loops.

\section{Summary}

We have presented a simple analytic proof of the NAST
for an unknotted loop. 
The validity of the NAST has been assured by obtaining explicit relations 
between loop variables and the Bianchi identity. We have shown that 
the Lagrangian density of the non-Abelian gauge field can be expressed 
by a set of six loop variables $W^{(\mu \nu)}(x)$, $\mu >\nu$.
We hope that the results obtained here are helpful for the discussion 
of the duality of the non-Abelian gauge field, where the deepest understanding
of the Bianchi identity will be indispensable.
In a future communication,
the NAST for knotted loops and links will be discussed.

\section*{Acknowledgements}
The authors are grateful to S. Hamamoto, T. Kurimoto,
H. Yamakoshi and M. Ueno for discussions and their interest.
Thanks are also due to M. B. Halpern, Yu. A. Simonov and V. I. Shevchenko
for comments.
The comment by J. S. Dowker that the NAST was for the first time
proposed by L. Schlesinger$^{14)}$ in 1927 should be especially
acknowledged.

\appendix
\section{Calculation of $\delta I[S]$}
By Eq. (3$\cdot$6), we have
\begin{eqnarray}
ig\frac{\delta I[S]}{\delta x^\rho (s,t)}=N_\rho (s,t)+M_\rho(s,t)
\end{eqnarray}
with
\begin{eqnarray}
N_\rho (s,t)&=&\int_{s_1}^{s_2}ds'\int_{t_1}^{t_2}dt' \hspace{1mm} v(s',t')
\frac{\delta K(s',t')}{\delta x^\rho (s,t)}v^{-1}(s',t'),\\
M_\rho (s,t)&=&\int_{s_1}^{s_2}ds'\int_{t_1}^{t_2}dt' \hspace{1mm} v(s',t')
\left[v^{-1}(s',t')\frac{\delta v(s',t')}{\delta x^\rho (s,t)},K(s',t')\right]
v^{-1}(s',t').\nonumber \\
\end{eqnarray}
In the following, we shall show that $N_\rho (s,t)$ and $M_\rho (s,t)$
are given by
\begin{eqnarray}
\frac{1}{ig}N_\rho (s,t)&=&v(s,t)\{P_\rho (s,t)
+Q_\rho (s,t)\}v^{-1}(s,t)+R_\rho (s,t),\\
\frac{1}{ig}M_\rho (s,t)&=&S_\rho (s,t)-v(s,t)Q_\rho (s,t)v^{-1}(s,t),
\end{eqnarray}
where $P_\rho$, $Q_\rho$, $R_\rho$ and $S_\rho$ are defined by
\begin{eqnarray}
P_\rho &=&x^\mu_s x^\nu_t ([D_\rho,F_{\mu \nu}(x)]
+[D_\nu,F_{\rho \mu}(x)]+[D_\mu,F_{\nu \rho}(x)]),\\
Q_\rho &=&[A_\rho(x),K],\\
R_\rho &=&x^\sigma_s\biggl[vF_{\rho \sigma}(x)v^{-1},
\int_{s_1}^{s}ds' \hspace{1mm} v(s',t)K(s',t)v^{-1}(s',t)\biggl],\\
S_\rho &=&x^\sigma_s\biggl[vF_{\rho \sigma}(x)v^{-1},
\int_{s}^{s_2}ds' \hspace{1mm} v(s',t)K(s',t)v^{-1}(s',t)\biggl]
\end{eqnarray}
with the abbreviated notation $P_\rho =P_\rho (s,t)$, 
$x^\mu_s=x^\mu_s(s,t)$, etc. We first discuss $N_\rho (s,t)$.
The functional derivative of $K(s',t')$ is readily calculated to be
\begin{eqnarray}
\frac{1}{ig}
\frac{\delta K(s',t')}{\delta x^\rho (s,t)}=\left( x^\mu_s x^\nu_t
(\partial_\rho F_{\mu \nu})+x^\nu_t F_{\rho \nu}
\frac{\partial}{\partial s'}+x^\mu_s F_{\mu \rho}
\frac{\partial}{\partial t'}\right) \delta(s-s')\delta(t-t').\nonumber \\
\end{eqnarray}
From (A$\cdot$2) and (A$\cdot$10), we obtain
\begin{eqnarray}
\frac{1}{ig}N_\rho=&x^\mu_s& x^\nu_tv(\partial_\rho F_{\mu \nu})v^{-1}
-\frac{\partial}{\partial s}(x^\nu_tvF_{\rho \nu}v^{-1})
-\frac{\partial}{\partial t}(x^\mu_svF_{\mu \rho}v^{-1}) \nonumber \\
=&v&(x^\mu_s x^\nu_t\partial_\rho F_{\mu \nu}-x^\mu_s x^\nu_t
\partial_\mu F_{\rho \nu}-x^\nu_t x^\mu_s
\partial_\nu F_{\mu \rho})v^{-1} \nonumber \\
&-&x^\nu_t 
v\left[v^{-1}\frac{\partial v}{\partial s},F_{\rho \nu}\right]v^{-1}
-x^\mu_s v\left[v^{-1}\frac{\partial v}{\partial t},F_{\mu \rho}\right]v^{-1}.
\end{eqnarray}
If we recall (2$\cdot$21), (2$\cdot$6), (2$\cdot$11) 
and (2$\cdot$15), all the necessary
differentiations are carried out, and we finally have (A$\cdot$4).

We now turn to the calculation of $M_\rho(s,t)$. Noting that $V(t_1,t';s_1)$
remains fixed under the variation $\delta x(s,t)$, we have
\begin{eqnarray}
\delta v(s',t')&=&\delta (V(t_1,t';s_1)U(s_1,s';t')) \nonumber \\
               &=&V(t_1,t';s_1) \delta U(s_1,s';t').
\end{eqnarray}
Since the variation $\delta U(s_1,s';t')$ consists of 
the deformation of the path $\{x(s'',t')|s_1$\\$< s''<s'\}$ and 
the displacement of the end point $x (s',t')$, we are led to
\begin{eqnarray}
\delta U(s_1,s';t')&=&\int_{s_1}^{s'}U(s_1,s'';t')\{
igF_{\mu \nu}(x(s'',t'))x^\mu_s(s'',t')ds''\delta x^\nu(s'',t')\}
U(s'',s';t') \nonumber \\
& &-igU(s_1,s';t')A_\mu(x(s',t'))\delta x^\mu(s',t').
\end{eqnarray}
The functional derivative $\frac{\delta v(s',t')}{\delta x (s,t)}$ is
then given by
\begin{eqnarray}
\frac{\delta v(s',t')}{\delta x^\rho (s,t)}&=&V(t_1,t';s_1)
\int_{s_1}^{s'}ds'' \hspace{1mm} U(s_1,s'';t')
igF_{\mu \rho}(x(s'',t'))x^\mu_s(s'',t') \nonumber \\
& &\times U(s'',s';t')\delta (s''-s)\delta (t'-t) \nonumber \\
& &-igv(s',t')A_\rho (x(s',t'))\delta (s'-s)\delta (t'-t),
\end{eqnarray}
yielding
\begin{eqnarray}
v^{-1}(s',t')\frac{\delta v(s',t')}{\delta x^\rho (s,t)}&=&
\delta (t'-t)\theta (s'-s)U(s',s;t)igF_{\mu \rho}(x (s,t))x^\mu_s(s,t)
U^{-1}(s',s;t) \nonumber \\
& &-igA_\rho (x(s,t))\delta (s'-s)\delta (t'-t),
\end{eqnarray}
with $\theta (s'-s)$ the step function. 
It is now easy to obtain (A$\cdot$5) from (A$\cdot$15) and (A$\cdot$3)
and see that (A$\cdot$1)-(A$\cdot$5) yield (3$\cdot$7)-(3$\cdot$10).

\section{Calculation of $P$}
From (A$\cdot$10) and (4$\cdot$9), we have
\begin{eqnarray}
P=ig\int_{s_1}^{s_2}&ds&\int_{t_1}^{t_2}dt \hspace{1mm} X(t)v(s,t) \nonumber \\
&\times& \int_{s_1}^{s_2}ds'\int_{t_1}^{t_2}dt' \hspace{1mm} 
\delta x^\rho (s',t') \nonumber \\
&\times& \left\{\left((\partial_\rho F_{\mu \nu})x^\mu_s x^\nu_t
+x^\nu_t F_{\rho \nu}
\frac{\partial}{\partial s}+x^\mu_s F_{\mu \rho}
\frac{\partial}{\partial t}\right)
\delta(s-s')\delta(t-t')\right\} \nonumber \\
&\times& v^{-1}(s,t)Y(t) \nonumber \\
=ig\int_{s_1}^{s_2}&ds'&\int_{t_1}^{t_2}dt' 
\hspace{1mm} \delta x^\rho (s',t') \nonumber \\
&\times&
\biggl[X(t')v(s',t')((\partial_\rho F_{\mu \nu})x^\mu_s x^\nu_t)(s',t')
v^{-1}(s',t')Y(t') \nonumber \\
& &\hspace{3mm}
-\{\partial_s (X(t)v(s,t)\{(F_{\rho \nu} x^\nu_t)(s,t)\}v^{-1}(s,t)Y(t))\}
_{s=s', t=t'} \nonumber \\
& &\hspace{3mm}
-\{\partial_t (X(t)v(s,t)\{(F_{\mu \rho} x^\mu_s)(s,t)\}v^{-1}(s,t)Y(t))\}
_{s=s', t=t'}\biggl].\nonumber \\
\end{eqnarray}
In (B$\cdot$1), we have made use of the relation
\begin{eqnarray}
\int_{s_1}^{s_2}&ds'&\int_{t_1}^{t_2}dt' \hspace{1mm} \delta x^\rho (s',t')
\int_{s_1}^{s_2}ds\int_{t_1}^{t_2}dt \hspace{1mm} g(s,t)
\delta(t-t')\frac{\partial}{\partial s}\delta(s-s') \nonumber \\
&=&\int_{s_1}^{s_2}ds'\int_{t_1}^{t_2}dt' \hspace{1mm} \delta x^\rho (s',t')
\int_{s_1}^{s_2}ds \hspace{1mm} g(s,t')\frac{\partial}{\partial s}
\delta(s-s') \nonumber \\
&=&\int_{s_1}^{s_2}ds'\int_{t_1}^{t_2}dt' \hspace{1mm} 
\delta x^\rho (s',t') \nonumber \\
&  &\times \left[g(s_2,t')\delta(s_2-s')-g(s_1,t')\delta(s_1-s')
-\int_{s_1}^{s_2}ds \hspace{1mm} \frac{\partial g(s,t')}{\partial s}
\delta(s-s')\right] \nonumber \\
&=&-\int_{s_1}^{s_2}ds'\int_{t_1}^{t_2}dt' \hspace{1mm} \delta x^\rho (s',t')
\frac{\partial g(s',t')}{\partial s'},
\end{eqnarray}
which results from the properties 
$\delta x^\rho (s_2,t')=\delta x^\rho (s_1,t')=0$.
The derivatives on the r.h.s. of (B$\cdot$1) are calculated 
with the aid of the following relations:
\begin{eqnarray}
\frac{\partial X(t)}{\partial s}&=&0,\\
\frac{\partial X(t)}{\partial t}&=&-X(t)B(t),\\
\frac{\partial Y(t)}{\partial s}&=&0,\\
\frac{\partial Y(t)}{\partial t}&=&B(t)Y(t).
\end{eqnarray}
We then have
\begin{eqnarray}
P&=&(-P_{23}-R)+(P_{21}+P_{22}+P_{23})  \nonumber \\
&=&P_{21}+P_{22}-R,
\end{eqnarray}
with $R$, $P_{21}$, $P_{22}$ and $P_{23}$ defined by
\begin{eqnarray}
R&=&ig\int_{s_1}^{s_2}ds\int_{t_1}^{t_2}dt \hspace{1mm} 
\delta x^\rho (s,t)X(t)v(s,t)  \nonumber \\
& &\hspace{10mm}
\times \left[\int_{s}^{s_2}ds' \hspace{1mm}U(s,s';t)K(s',t)U(s',s;t), 
(F_{\mu \rho}x^\mu_s)(s,t)\right]  \nonumber \\
& &\hspace{10mm}\times v^{-1}(s,t)Y(t),\\
P_{21}&=&ig\int_{s_1}^{s_2}ds\int_{t_1}^{t_2}dt \hspace{1mm} 
\delta x^\rho (s,t)X(t)v(s,t)P_\rho (s,t)v^{-1}(s,t)Y(t),\\
P_{22}&=&ig\int_{s_1}^{s_2}ds\int_{t_1}^{t_2}dt \hspace{1mm} 
\delta x^\rho (s,t)X(t)v(s,t)Q_\rho (s,t)v^{-1}(s,t)Y(t),\\
P_{23}&=&-ig\int_{s_1}^{s_2}ds\int_{t_1}^{t_2}dt \hspace{1mm} 
\delta x^\rho (s,t)X(t)R_\rho (s,t)Y(t).
\end{eqnarray}
The functions $P_\rho$, $Q_\rho$ and $R_\rho$ in (B$\cdot$9)-(B$\cdot$11)
 are those 
given in (A$\cdot$6), (A$\cdot$7) and (A$\cdot$8), respectively.

\section{Calculation of $Q$}
The variation $\delta U(s_1,s;t)$ is given by (A$\cdot$13).
The functional derivative $\frac{\delta U(s_1,s;t)}{\delta x^\rho (s',t')}$
is now calculated to be
\begin{eqnarray}
\frac{\delta U(s_1,s;t)}{\delta x^\rho (s',t')}&=&
U(s_1,s';t)igF_{\mu \rho}(x(s',t))x^\mu_s(s',t)U(s',s;t)\theta (s-s')
  \delta (t-t') \nonumber \\
& &-U(s_1,s;t)igA_\rho (x(s,t))\delta (s-s')\delta (t-t'),
\end{eqnarray}
where $\theta$ is the step function. From the above, 
we see that $Q$ in (4$\cdot$10) is given by
\begin{eqnarray}
Q&=&\int_{s_1}^{s_2}ds\int_{t_1}^{t_2}dt \hspace{1mm} X(t)v(s,t)
  \int_{s_1}^{s_2}ds'\int_{t_1}^{t_2}dt' \hspace{1mm} \delta x^\rho(s',t')
\delta(t-t') \nonumber \\
& &\times
[U(s,s';t)(igF_{\mu \rho}x^\mu_s)(s',t)U(s',s;t)\theta(s-s')
-igA_\rho(x(s,t))\delta(s-s'), K(s,t)] \nonumber \\
& &\times v^{-1}(s,t)Y(t) \nonumber \\
&=&\int_{s_1}^{s_2}ds'\int_{t_1}^{t_2}dt' \hspace{1mm} 
\delta x^\rho(s',t') X(t')\int_{s'}^{s_2}ds \hspace{1mm} \nonumber \\
& &\times [v(s',t')\{-ig(F_{\rho \sigma}x^\sigma_s)(s',t')\}v^{-1}(s',t'),
v(s,t')K(s,t')v^{-1}(s,t')]Y(t') \nonumber \\
& &-P_{22},
\end{eqnarray}
where we have made use of the equality $v(s,t')U(s,s';t')=v(s',t')$,
 and $P_{22}$ is 
the quantity defined in (B$\cdot$10). The first term on the r.h.s. 
of (C$\cdot$2) is calculated as follows:
\begin{eqnarray}
Q+P_{22}&=&-ig\int_{s_1}^{s_2}ds\int_{t_1}^{t_2}dt \hspace{1mm} 
\delta x^\rho(s,t)X(t)\int_{s}^{s_2}ds' \nonumber \\
& &\hspace{1.5cm}\times 
[v(s,t)(F_{\rho \sigma}x^\sigma_s)(s,t)v^{-1}(s,t),
v(s',t)K(s',t)v^{-1}(s',t)]Y(t) \nonumber \\
&=&ig\int_{s_1}^{s_2}ds\int_{t_1}^{t_2}dt \hspace{1mm}
\delta x^\rho(s,t)X(t)v(s,t)\int_{s}^{s_2}ds' \nonumber \\
& &\hspace{1.5cm}\times [U(s,s';t)K(s',t)U^{-1}(s,s';t), 
(F_{\rho \sigma}x^\sigma_s)(s,t)]v^{-1}(s,t)Y(t), \nonumber \\
\end{eqnarray}
where use has been made of the relation $v^{-1}(s,t)v(s',t)=U(s,s';t)$.
Since the last expression coincides with the r.h.s. of (B$\cdot$8),
we conclude (4$\cdot$12).

\section{Calculation of $T_{uv}-T_{vu}$}
Combining (5$\cdot$9) with (5$\cdot$2), we obtain
\begin{eqnarray}
T_{uv}&=&\left[j_u(0),\frac{\partial W[\gamma]}{\partial v}\right]+
\left[j_v(0),\frac{\partial W[\gamma]}{\partial u}\right]
+j_u(0)W[\gamma]j_v(0)+j_v(0)W[\gamma]j_u(0) \nonumber \\
& &\hspace{1mm}+\left[\frac{\partial j_v(0)}{\partial u},W[\gamma]\right]
-j_u(0)j_v(0)W[\gamma]-W[\gamma]j_v(0)j_u(0)  \nonumber \\
& &\hspace{1mm}-\int_0^4dr \hspace{1mm}\omega (4,r)\left( [K_v(r),j_u(r)]
+\frac{\partial K_v(r)}{\partial u}\right)
\omega(r,0).
\end{eqnarray}
We then have
\begin{eqnarray}
T_{uv}-T_{vu}&=&\left[\frac{\partial j_v(0)}{\partial u}
-\frac{\partial j_u(0)}{\partial v}
-[j_u(0),j_v(0)],W[\gamma]\right] \nonumber \\
& &\hspace{1mm}-\int_0^4dr \hspace{1mm} \omega (4,r)
\biggl( \frac{\partial K_v(r)}{\partial u}
-\frac{\partial K_u(r)}{\partial v} \nonumber \\
& &\hspace{3.3cm}-[j_u(r),K_v(r)]+[j_v(r),K_u(r)]\biggl)
\omega(r,0) \nonumber \\
&=&ig[x^\mu_u(0)x^\nu_v(0)F_{\mu \nu}(x(0)),W[\gamma]] \nonumber \\
& &\hspace{1mm}-ig\int_0^4dr \hspace{1mm} 
\omega (4,r)x^\mu_r(r) \nonumber \\
& &\hspace{1.5cm}\times \{x^\nu_v(r)x^\rho_u(r)
[D_\rho, F_{\mu \nu}(x(r))]-x^\nu_u(r)x^\rho_v(r)
[D_\rho, F_{\mu \nu}(x(r))]\}\nonumber \\
& &\hspace{1.5cm}\times \omega(r,0) \nonumber \\
& &\hspace{1mm}-ig\int_0^4dr 
\hspace{1mm} \omega (4,r)\left\{\frac{\partial}{\partial r}
(x^\mu_u(r)x^\nu_v(r))\right\}F_{\mu \nu}(x(r))\omega(r,0),
\end{eqnarray}
where we have assumed $\partial_u\partial_v x^\mu(r)
=\partial_v\partial_u x^\mu(r)$. Through a partial integration,
it can be seen that the last term on the r.h.s. of (D$\cdot$2) is equal to
\begin{eqnarray}
-ig[&x^\mu_u&(0)x^\nu_v(0)F_{\mu \nu}(x(0)),W[\gamma]] \nonumber \\
&+&ig\int_0^4dr \hspace{1mm} x^\rho_u(r)x^\nu_v(r)x^\mu_r(r)
\omega (4,r)[D_\mu, F_{\rho \nu}(x(r))]\omega (r,0).
\end{eqnarray}
These observations lead us to (5$\cdot$14).

\end{document}